\documentclass[twocolumn,showpacs,preprintnumbers,amsmath,amssymb]{revtex4}

\usepackage{graphicx}
\usepackage{dcolumn}
\usepackage{bm}
\usepackage{amssymb}
\usepackage{indentfirst}
\usepackage{psfig,color}
\usepackage{epsfig}
\usepackage{epsf}
\usepackage{graphicx}

\def\Vec#1{{\bf #1}}

\begin{document}

\preprint{USM-TH-180}

\title{Pion pair production in $e^+e^-$
annihilation}
\author{Zhun Lu}\email[Electronic address: ]{zhun.lu@usm.cl}
\author{Ivan
Schmidt}\email[Electronic address:
]{ivan.schmidt@usm.cl}\affiliation{Departamento de F\'\i sica,
Universidad T\'ecnica Federico Santa Mar\'\i a, Casilla 110-V,
Valpara\'\i so, Chile}

\begin{abstract}
We present an analysis of the process $e^+e^-\rightarrow \gamma^*\,
\textrm{or} \, Z^*\rightarrow\pi\pi\gamma$, in the kinematical
region where $\sqrt{s}$, the c.m. energy of the $e^+e^-$ pair, is
large but much below the $Z$-pole. The subprocess $\gamma^*\,
\textrm{or} \, Z^*\rightarrow \pi\pi \gamma$ can be described by the
convolution of the hard scattering coefficient $\gamma^*\,
\textrm{or} \, Z^*\rightarrow q \bar{q} \gamma$ and the general
distribution amplitude of two pions $q\bar{q}\rightarrow \pi \pi$.
In the case of neutral pion, the production through $\gamma^*$ is
the dominant process, which can therefore be used to access the
generalized distribution amplitudes (GDAs) of the pion, especially
their $C$-even parts. The $\gamma Z$ interference term provides an
alternative approach to extract the weak mixing angle $\sin
\theta_W$ through measuring the helicity asymmetry in the process
$e^+e^-\rightarrow \pi^0\pi^0\gamma$. In the case of charged pion
pair production, the Bremsstrahlung process dominates and its
interference with $e^+e^-\rightarrow \gamma^ * \rightarrow
\pi^+\pi^-\gamma$ can be applied to study the process $\gamma^ *
\rightarrow \pi\pi\gamma$ at the amplitude level.
\end{abstract}


\maketitle

\section{Introduction}

Generalized distribution amplitudes (GDAs) \cite{Diehl98,muller94}
are important nonperturbative objects for understanding how quarks
and gluons form hadrons in hard exclusive processes. They describe
the soft transition $q \bar{q} \rightarrow h \bar{h}$ or $g g
\rightarrow h \bar{h}$, in the kinematical regime where the
invariant mass of the hadron pair $h\bar{h}$ is small compared to
the hard scale of the process. The typical processes to which the
GDAs formalism can be applied are the two photon process
$\gamma^*\gamma \rightarrow h \bar{h}$~\cite{Diehl00,dkv03,apt04}
and hadron pair eletroproduction $\gamma^* p\rightarrow p^\prime h
\bar{h}~$\cite{Polyakov99,lppsg}, for which factorization theorems
have been proved~\cite{freund00}. In Ref.~\cite{ls05} we gave an
analysis of the process $e^+e^-\rightarrow Z \rightarrow \pi \pi
\gamma$ at the $Z$-pole energy, with small pion pair invariant mass.
The subprocess $Z \rightarrow \pi \pi \gamma$ can be factorized into
the convolution of a hard coefficient and the generalized
distribution amplitude of two pions (2-pion GDA). In the case of
charged pion pair production, the process is sensitive to the
$C$-odd parts of 2-pion GDAs, and provide an opportunity to access
them in $e^+e^-$ annihilation. In this work we extend the analysis
in Ref.~\cite{ls05} to a generalized process $e^+e^-\rightarrow
\gamma^*\, \textrm{or} \, Z^*\rightarrow\pi\pi\gamma$, in the
kinematical region where $\sqrt{s}$ is large but much below the
$Z$-pole. Like the case of the production through $Z$ boson analyzed
in Ref.~\cite{ls05}, a similar factorization holds for the case of
the production through a virtual photon, which gives the dominant
contribution for neutral pion pair production in the kinematical
region we discuss here. Therefore, the investigation of the process
$e^+e^-\rightarrow \gamma^*\rightarrow\pi^0\pi^0\gamma$ can provide
detailed information about 2-pion GDAs, especially their $C$-even
part. We also discuss the $\gamma Z$ interference term, which
contains the weak mixing angle $\sin \theta_W$ and can be used as an
alternative approach to extract its value. The Bremsstrahlung
process $e^+e^-\rightarrow\gamma^*\gamma\rightarrow\pi^+\pi^-\gamma$
dominates in the case of charged pion production, and its
interference with $e^+e^-\rightarrow \gamma^ * \rightarrow
\pi^+\pi^-\gamma$ is a valuable tool for studying the process
$\gamma^ * \rightarrow \pi\pi\gamma$ at the amplitude level.

\section{Analysis of the process
$\gamma^*\rightarrow\pi\pi\gamma$}\label{gammagamma}

\begin{figure}

\begin{center}
\scalebox{1.05}{\includegraphics{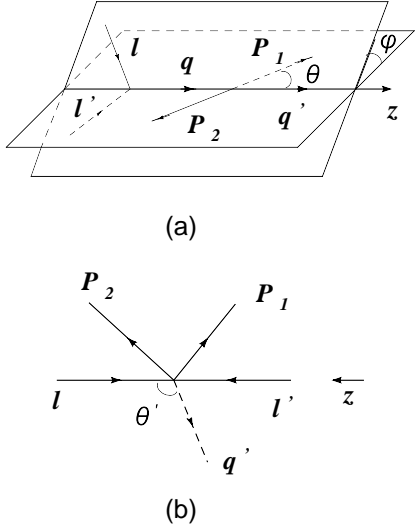}} \caption{\small The
kinematics of pion pair production. (a) the c.m. frame of the pion
pair, (b) the c.m. frame of $e^+e^-$. } \label{kine}
\end{center}

\end{figure}

We first study the subprocess of pion pair production
\begin{equation}
\gamma^*(q)\rightarrow \pi(p_1)+\pi(p_2)+
\gamma(q^\prime)\label{gamma}
\end{equation}
in the framework of GDAs. It is convenient to investigate this
subprocess in the pion pair center of mass frame (shown in
Fig.~\ref{kine}a), where the kinematics is expressed as (choosing
the $z$ axis along the momentum of the real photon)
\begin{eqnarray}
q&=&\frac{Q}{\sqrt{2}}v+\frac{Q}{\sqrt{2}}v^\prime,~~~
q^\prime=\frac{Q^2-W^2}{\sqrt{2}Q}v^\prime,\nonumber \\
P&=&\frac{Q}{\sqrt{2}}v+\frac{W^2}{\sqrt{2}Q}v^\prime,
\nonumber \\
p_1&=&\frac{\zeta Q}{\sqrt{2}}v+\frac{(1-\zeta)W^2}
{\sqrt{2}Q}v^{\prime}+\frac{\Delta_T}{2},\nonumber\\
p_2&=&\frac{(1-\zeta)Q}{\sqrt{2}}v+\frac{\zeta
W^2}{\sqrt{2}Q}v^{\prime}-\frac{\Delta_T}{2}
\end{eqnarray}
respectively. Here $v$ and $v^\prime$ are two lightlike vectors
which satisfy: $v^2=v^{\prime 2}=0$, $v\cdot v^\prime=1$, $W$ is the
invariant mass of the pion pair and $q^2=Q^2=s$, $P$ is the total
momentum of the pion pair. The skewness parameter $\zeta$ is the
momentum fraction of plus momentum carried by $\pi(p_1)$ with
respect to the pion pair:
\begin{equation}
\zeta=\frac{p_1^+}{P^+}=\frac{1+\beta \cos
\theta}{2},~~~~\beta=\sqrt{1-\frac{4m_\pi^2}{W^2}}.
\end{equation}

\begin{figure*}

\begin{center}
\scalebox{0.88}{\hspace{-0.5cm}\includegraphics{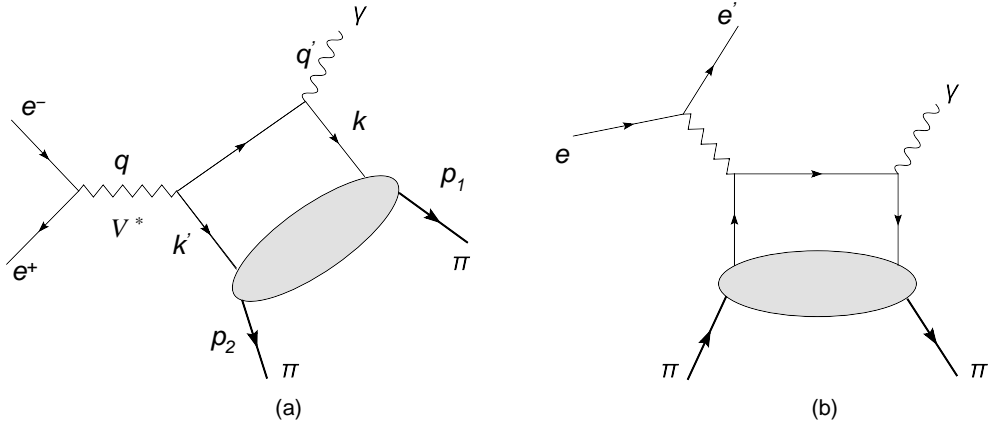}}
\caption{\small (a) Diagram of the process $e^+e^-\rightarrow V^*
\rightarrow \pi \pi \gamma$, where $V$ can be a photon or $Z$ boson.
(b) Diagram of the DVCS process $e\pi\rightarrow  e \pi \gamma$, the
crossing related channel of the process depicted in (a).}
\label{pionpair}
\end{center}

\end{figure*}

In the kinematical region $W^2 \ll s$, the process given in
Eq.~(\ref{gamma}) can be analyzed in the framework of GDAs, which
has been applied in Ref.~\cite{ls05} to the case where the virtual
photon is replaced by the $Z$ boson at $\sqrt{s}=M_Z$. From
Fig.~\ref{pionpair}a and the one with the two photon vertices
interchanged, the hadronic tensor of the process can be calculated
as~\cite{Diehl98,ls05}
\begin{eqnarray}
T^{\mu\nu}_{\gamma^*\rightarrow \pi\pi\gamma}&=&i\int
d^4xe^{-iq\cdot x}\langle
\pi(p_1)\pi(p_2)|TJ_{em}^\mu(x)J_{em}^\nu(0)|0\rangle
\nonumber\\
&=&-g^{\mu\nu}_\perp\sum_q \frac{e^2 e^2_q}{2} \int_0^1 dz
\frac{2z-1}{z(1-z)}\Phi_q^{\pi\pi}(z,\zeta,W^2),\nonumber\\\label{gammatensor}
\end{eqnarray}
which has the same form as that of the two photon
process~\cite{Diehl98,Diehl00}, and $\Phi_q(z,\zeta,W^2)$ is the GDA
defined as~\cite{Diehl98}:
\begin{eqnarray}
\Phi_q(z,\zeta,W^2)&=&\int \frac{dx^-}{2\pi} e^{iz(P^+
x^-)}\nonumber\\
&&\times\langle
\pi(p_1)\pi(p_2)|\bar{\psi}(x^-)\gamma^+\psi(0)|0\rangle,
\end{eqnarray}
where $z=k^+/P^+$ is the momentum fraction of the quark with
respect to the hadronic system.

As discussed in Refs.~\cite{Diehl98,Polyakov99}, a useful constraint
on the GDAs comes from charge conjugation invariance of the strong
interactions:
\begin{equation}
\Phi_q^{\pi\pi}(z,\zeta,W^2)=-\Phi_q^{\pi\pi}(1-z,1-\zeta,W^2).
\end{equation}
Following this constraint one can define the $C$-even and $C$-odd
parts (in \cite{Polyakov99} they are also called isoscalar and
isovector parts respectively) of 2-pion GDA as~\cite{Diehl00}
\begin{equation}
\Phi_q^{\pm}(z,\zeta,W^2)=\frac{1}{2}[\Phi_q^{\pi\pi}(z,\zeta,W^2)
\pm \Phi_q^{\pi\pi}(z,1-\zeta,W^2)],\label{cpm}
\end{equation}
where the superscript $\pi \pi$ represents $\pi^0 \pi^0$ or $\pi^+
\pi^-$. Then
\begin{equation}
\Phi_q^{\pi\pi}(z,\zeta,W^2)=\Phi_q^{+}(z,\zeta,W^2)+\Phi_q^{-}(z,\zeta,W^2).\label{cdecomp}
\end{equation}
The properties of $\Phi_q^{\pm}(z,\zeta,W^2)$ under the interchanges
$z \rightarrow 1-z$ and $\zeta \rightarrow 1-\zeta$ can be easily
derived from (\ref{cdecomp})~\cite{Polyakov99}:
\begin{eqnarray}
\Phi_q^{+}(z,\zeta,W^2)&=&-\Phi_q^{+}(1-z,\zeta,W^2)\nonumber\\
&=&\Phi_q^{+}(z,1-\zeta,W^2),\label{ceven}\\
\Phi_q^{-}(z,\zeta,W^2)&=&\Phi_q^{-}(1-z,\zeta,W^2)\nonumber\\
&=&-\Phi_q^{-}(z,1-\zeta,W^2).\label{codd}
\end{eqnarray}
The $\zeta \rightarrow 1-\zeta$ exchange corresponds to the
interchange of the two pions, thus in the case in which the final
two pions are $\pi^0 \pi^0$ we get
\begin{equation}
\Phi_q^{\pi^0 \pi^0}(z,\zeta,W^2)=\Phi_q^{\pi^0
\pi^0}(z,1-\zeta,W^2).
\end{equation}
Therefore there is only a $C$-even part for $\Phi_q^{\pi^0
\pi^0}(z,\zeta,W^2)$. In the following we use
$\Phi_q^{\pm}(z,\zeta,W^2)$ to represent the $C$-even/odd part of
$\Phi_q^{\pi^+\pi^-}(z,\zeta,W^2)$ respectively. Isospin invariance
implies~\cite{Diehl00}
\begin{eqnarray}
\Phi_q^{\pi^0\pi^0}(z,\zeta,W^2)=\Phi_q^{+}(z,\zeta,W^2).
\end{eqnarray}
Thus according to (\ref{gammatensor}), only
$\Phi_q^{+}(z,\zeta,W^2)$ contributes to the process
$\gamma^*\rightarrow \pi\pi\gamma$. Here we only consider the
contributions coming from the $u$ and $d$ quark. Another consequence
of isospin invariance is
\begin{eqnarray}
\Phi^{+}_u=\Phi^{+}_d,~~~~\Phi^{-}_u=-\Phi^{-}_d.
\end{eqnarray}

Eq.~(\ref{gammatensor}) shows that the process $e^+e^- \rightarrow
\gamma^* \rightarrow \pi\pi\gamma$ can be also applied to probe
$\Phi^{+}$, just like the two photon process $\gamma^* \gamma
\rightarrow \pi \pi$, which has been proposed~\cite{Diehl00} to
access $\Phi^{+}$. Interest on the process $e^+e^- \rightarrow
\gamma^* \rightarrow \pi\pi\gamma$ relies also on the fact that this
process is related to its crossed channel, the deeply virtual
Compton scattering process (DVCS) $e \pi \rightarrow e\pi\gamma$
(shown in Fig.~\ref{pionpair}b), and is referred to as timelike
DVCS. In the factorization of the DVCS process, the nonperturbative
objects are the generalized parton distributions (GPDs)
\cite{dm,dvcs}, which are important for understanding the parton
angular momenta inside the nucleon~\cite{ji97}. The simplest DVCS
process is $e \pi \rightarrow e\pi\gamma$. However it is difficult
to use the pion as the target, which makes the pion GPDs difficult
to be experimentally accessed. It has been shown~\cite{teryaev} that
there is a crossing relation that connects GPDs and GDAs. Therefore
to measure the process $e^+e^- \rightarrow \gamma^* \rightarrow
\pi\pi\gamma$, the crossed channel of the DVCS process, can also
give experimental indication about the GPDs of the pion. In the
kinematical region $W^2\ll s \ll M_Z^2$, the suitable facilities to
investigate this process are $B$-factories. Finally, we remark that
there is a suggestion~\cite{ps05b} of using the transition
distribution amplitude (TDA)~\cite{ps05} formalism to describe the
process $e^+e^-\rightarrow\pi\pi\gamma$.

\section{Phenomenology about the $e^+e^-\rightarrow \pi\pi\gamma$ process}

\subsection{Calculation of the cross-section}
 In this section we study the phenomenology associated
with the process \begin{equation}
 e^+(l)+e^-(l^\prime) \rightarrow V^*(q) \rightarrow
\pi(p_1)+\pi(p_2)+\gamma(q^\prime),
\end{equation}
where the intermediate vector boson $V$ can be a photon or a $Z$
boson, based on the analysis given in Section \ref{gammagamma} and
in Ref.~\cite{ls05}. We will consider the helicity of the electron
in the analysis. It is useful to define two Lorentz invariant
variables:
\begin{equation}
x=\frac{2q\cdot q^\prime}{q^2}, \;\;\;\; y=\frac{l\cdot
q^\prime}{q\cdot q^\prime}.
\end{equation}
 The
differential cross-section for the process $e^+e^-\rightarrow
V^*\rightarrow\gamma\pi\pi$ is expressed as
\begin{eqnarray}
d\sigma_{e^+e^-\rightarrow V^* \rightarrow\gamma\pi\pi}
&=&\frac{\beta
W(s-W^2)}{64(2\pi)^5s^2}|\mathcal{A}_{e^+e^-\rightarrow
V^* \rightarrow \pi\pi\gamma}|^2\nonumber\\
&&\times dWd\Omega^* d\Omega^\prime,\label{csgandz}
\end{eqnarray}
in which ($|\Vec p^*|,\Omega^*$) is the momentum of pion 1 in the
c.m. frame of the pion pair, and $\Omega^\prime$ is the angle of the
photon in the rest frame of the $Z$ boson (that is, the c.m. frame
of $e^+e^-$).

The amplitude of the process can be calculated from
\begin{eqnarray}
&&\mathcal{A}_{e^+e^-\rightarrow V^* \rightarrow
\gamma\pi\pi}=\sum_{i,j}\bar{v}(l^\prime)V^\mu
u(l) \epsilon^{(i)*}_\mu\nonumber\\
&&\hspace{1cm}\times\frac{1}{s-M_V^2}\mathcal{A}_{i,j}^{V^*\rightarrow
\gamma\pi\pi},
\end{eqnarray}
where $V^\mu=e\gamma^\mu(1+\lambda_e\gamma_5)$ for lepton-$\gamma$
vertex, and
$V^\mu=e\gamma^\mu(c^l_V-\gamma_5c^l_A)(1+\lambda_e\gamma_5)/\sin
2\theta_W$ for lepton-$Z$ boson vertex, with $\lambda_e=\pm 1$ the
helicity of the electron. The amplitude of $V^*\rightarrow
\pi\pi\gamma$ can be calculated from
\begin{equation}
\mathcal{A}_{i,j}^{V^*\rightarrow
\gamma\pi\pi}=\epsilon^{(i)}_\alpha\epsilon^{\prime (j)*}_\beta
T^{\alpha\beta}_V(z,\zeta,W^2),
\end{equation}
where $\epsilon$ and $\epsilon^\prime$ are the polarization vectors
of the intermediate virtual boson (photon or $Z$) and the real
photon, respectively. In our reference frame these vectors have the
form:
\begin{eqnarray}
\epsilon^{(\pm)}_\mu&=&\left(0,\frac{\mp
1}{\sqrt{2}},\frac{-i}{\sqrt{2}},0\right
),\epsilon_\mu^0=\left(\frac{|\Vec
q|}{\sqrt{s}},0,0,\frac{q_0}{\sqrt{s}}\right ),\nonumber\\
\epsilon^{\prime(\pm)}_\mu&=&\left(0,\frac{\mp
1}{\sqrt{2}},\frac{-i}{\sqrt{2}},0\right ).
\end{eqnarray}

Using the notation of Ref.~\cite{bjm99} the cross-section of the
process can be written as:
\begin{eqnarray}
\frac{d\sigma}{dW^2 d\cos \theta d\phi dy}=\frac{\alpha^3 \beta
(s-W^2)}{32\pi
s^4}\sum_{ab}L_{\mu\nu}^{ab}\mathcal{W}_{ab}^{\mu\nu}\chi_{ab}.\label{csgz}
\end{eqnarray}
The indices $a$, $b$ can be $\gamma$ for the photon or $Z$ for the
$Z$ boson. The relative propagator factors $\chi_{ab}$ are given
by
\begin{eqnarray}
\chi_{\gamma \gamma}&=&1,\label{gg}\\
\chi_{\gamma
Z}&=&\chi_{Z\gamma}=\frac{1}{\sin^2(2 \theta_W)}\frac{s}{s-M_Z^2},\label{gz}\\
\chi_{ZZ}&=&(\chi_{\gamma Z})^2.\label{zz}
\end{eqnarray}
which correspond to the contributions from $\gamma$, $\gamma Z$
interference and $Z$ boson. The lepton tensors are:
\begin{eqnarray}
L_{ab}^{\mu\nu}=-2sC_{ab}I_1(y)g_\perp^{\mu\nu}-isD_{ab}I_2(y)\epsilon_\perp^{\mu\nu}.
\label{ltensor}
\end{eqnarray}

\begin{figure}

\begin{center}
\scalebox{0.9}{\includegraphics{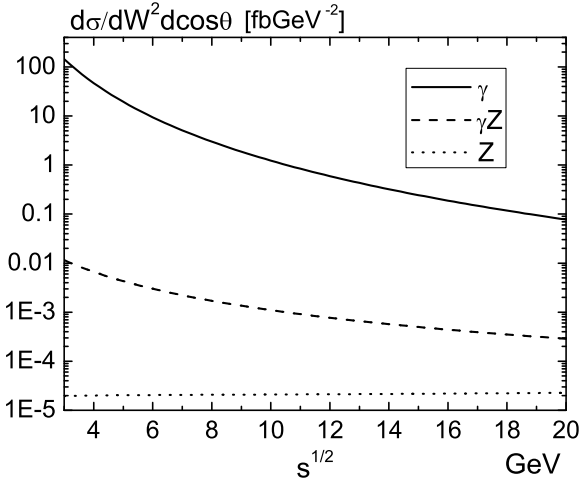}} \caption{\small The
$\gamma$ (solid line), $\gamma Z$ interference term (dashed line)
and $Z$ (dotted line) contributions to the helicity-independent
differential cross-section of the process $e^+e^-\rightarrow V^*
\rightarrow \pi^0\pi^0\gamma$ at $W=0.7\,\textrm{GeV}$,
$\theta=10^\circ$.} \label{csection}
\end{center}

\end{figure}

The notation used in the above equation is
\begin{eqnarray}
C_{\gamma\gamma}&=&1,\\
 C_{\gamma_Z}&=&C_{Z\gamma}=(c^l_V-c^l_A\lambda_e),\\
 C_{ZZ}&=&(c^{l2}_V+c^{l2}_A)-2c^l_Vc^l_A\lambda_e,\\
 D_{\gamma\gamma}&=&\lambda_e,\\
 D_{\gamma_Z}&=&D_{Z\gamma}=(c^l_V\lambda_e-c^l_A),\\
 D_{ZZ}&=&(c^{l2}_V+c^{l2}_A)\lambda_e-2c^l_Vc^l_A,\\
I_1(y)&=&(\frac{1}{2}-y+y^2),\\
I_2(y)&=&(1-2y).
\end{eqnarray}

 The hadron tensors appearing in Eq.~(\ref{csgz}) are:
\begin{eqnarray}
W^{\mu\nu}_{\gamma\gamma}&=& -g_\perp^{\mu
\nu}\left | \sum_qe_q^2V_q(\cos\theta,W^2)\right |^2, \label{wgg} \\
 W^{\mu\nu}_{\gamma Z}&=& -g_\perp^{\mu
\nu}\textrm{Re}\left \{\left [ \sum_q e_q c_V^q V_q^*(\cos\theta,W^2)\right ]\right.\nonumber\\
&&\left. \times \left [ \sum_q e_q^2 V_q(\cos\theta,W^2)\right ] \right\}\nonumber\\
&&+i\epsilon_\perp^{\mu
\nu}\textrm{Im}\left \{\left [ \sum_q e_q a_V^q V_q^*(\cos\theta,W^2)\right ]\right.\nonumber\\
&&\left. \times \left [ \sum_q e_q^2 A_q(\cos\theta,W^2)\right ] \right\}, \label{wgz} \\
W^{\mu\nu}_{Z Z}&=& -g_\perp^{\mu \nu} \left \{ \left | \sum_q e_q
c_V^q
V_q(\cos\theta,W^2)\right |^2 \right. \nonumber\\
&&+\left .\left | \sum_q e_q c_A^q A_q(\cos\theta,W^2)\right |^2 \right \} \nonumber\\
&&-2i\epsilon_\perp^{\mu\nu}\textrm{Im}\left \{\left [ \sum_q e_q
c_V^q V_q^*(\cos\theta,W^2)\right ]\right.\nonumber
\\
&& \times \left .\left [ \sum_q e_q c_A^q
A_q(\cos\theta,W^2)\right ]\right \},
\end{eqnarray}
where
\begin{eqnarray}
\hspace{-0.5cm}V_q(\cos\theta,W^2)&=& \int_{0}^{1}
dz\frac{2z-1}{z(1-z)}\Phi_q^+(z,\zeta,W^2),\\
\hspace{-0.5cm}A_q(\cos\theta,W^2)&=& \int_{0}^{1}
dz\frac{1}{z(1-z)}\Phi_q^-(z,\zeta,W^2).
\end{eqnarray}
The vector and axial-vector couplings to the $Z$ boson appearing in
the above equations are given by:
\begin{eqnarray}
c^i_V &=& T_3^i-2Q^i\sin^2\theta_W, \label{cqv}\\
c^i_A &=& T_3^i,\label{cqa}
\end{eqnarray}
where $Q^i$ denotes the charge and $T_3^i$ is the weak isospin of
particle $i$ (i.e., $T_3^i=+1/2$ for $i=u$, and $-1/2$ for
$i=e^-$, $d$, $s$).

\subsection{Results for the neutral case}

In the previous subsection we have given the expression for the
cross-section of the general process $e^+e^-\rightarrow
\gamma^*\,\textrm{or} \,Z^* \rightarrow \pi \pi \gamma$, in terms of
2-pion GDAs. There are interesting phenomenological implications
coming from Eq.~(\ref{csgandz}). As demonstrated in
Ref.~\cite{ls05}, measuring the charged pion pair production in the
kinematical regime $\sqrt{s}=M_Z$, where the main contribution comes
from $Z$ boson, can provide an opportunity to access the $C$-odd
part of 2-pion GDAs $\Phi^-_q$. In this work, however, we consider a
different kinematical region $W^2 \ll s\ll M_Z^2$. Consequently a
different phenomenology emerges: the production through $\gamma^*$
is dominant over the $\gamma Z$ interference and $Z$ boson terms,
which is a result coming from the factors given in Eqs.~(\ref{gg}),
(\ref{gz}) and (\ref{zz}).

To avoid the competing contribution from the Bremsstrahlung process,
in this subsection we limit ourselves to the case of neutral pion
pair production. Therefore the terms containing $A_q$ in the
hadronic tensor (\ref{ltensor}) do not contribute to the
cross-section, and the calculation is simplified. Based on
(\ref{csgandz}), we calculate the $\gamma$, $\gamma Z$ interference
and $Z$ contributions to the unpolarized differential cross-section
of the process $e^+e^-\rightarrow \gamma^* \,\textrm{or}\,Z^*
\rightarrow \pi^0\pi^0\gamma$ as functions of $\sqrt{s}$. The
results are shown in Fig.~\ref{csection} by a solid line, dashed
line and dotted line, respectively. For $\Phi_q^+$, which is needed
in the calculation, we adopt the simple model given in
\cite{Diehl00}. The curves show that the contribution from the
production through $\gamma^*$ is about several orders of magnitude
larger than those through $\gamma Z$ interference and $Z$ boson, in
the regime $W^2 \ll s \ll M_Z^2$. Therefore it would be suitable to
investigate $\Phi^+_q$ through the process $e^+e^- \rightarrow
\gamma^* \rightarrow \pi^0\pi^0 \gamma$, in the regime where
$\sqrt{s}$ is several GeV, where the cross-section is relatively
large.

\begin{figure}

\begin{center}
\scalebox{0.9}{\includegraphics{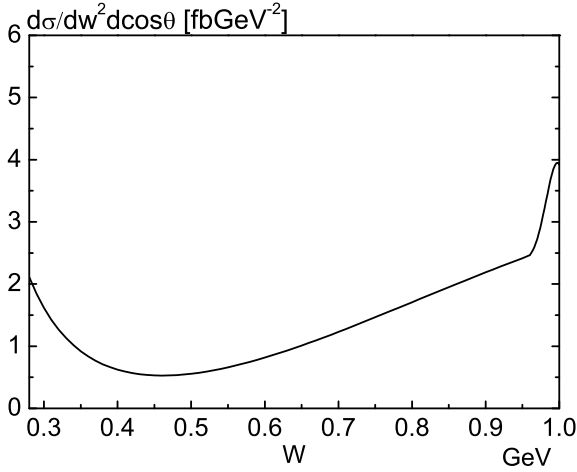}} \caption{\small The
$W$-dependent differential cross-section of the process
$e^+e^-\rightarrow \gamma^* \rightarrow \pi^0\pi^0\gamma$ at
$\sqrt{s}=10$ GeV and $\theta=20^\circ$.} \label{neutralw}
\end{center}

\end{figure}

\begin{figure}

\begin{center}
\scalebox{0.9}{\includegraphics{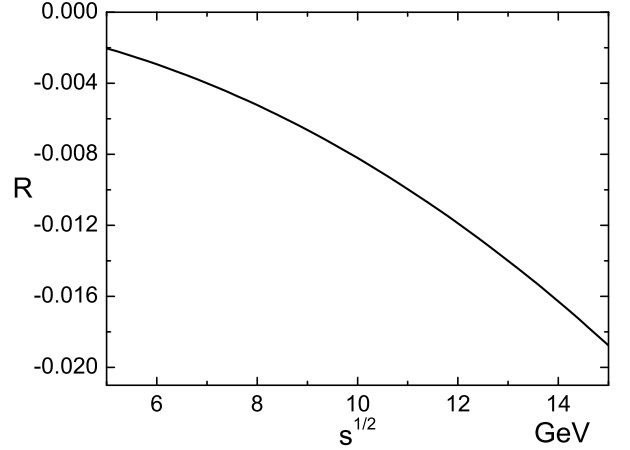}} \caption{\small The
$\sqrt{s}$ dependent ratio of the helicity dependent and
helicity-independent cross-section of the process $e^+e^-\rightarrow
V^* \rightarrow \pi^0\pi^0\gamma$.} \label{rs}
\end{center}

\end{figure}

\begin{figure*}

\begin{center}
\scalebox{1.3}{\includegraphics{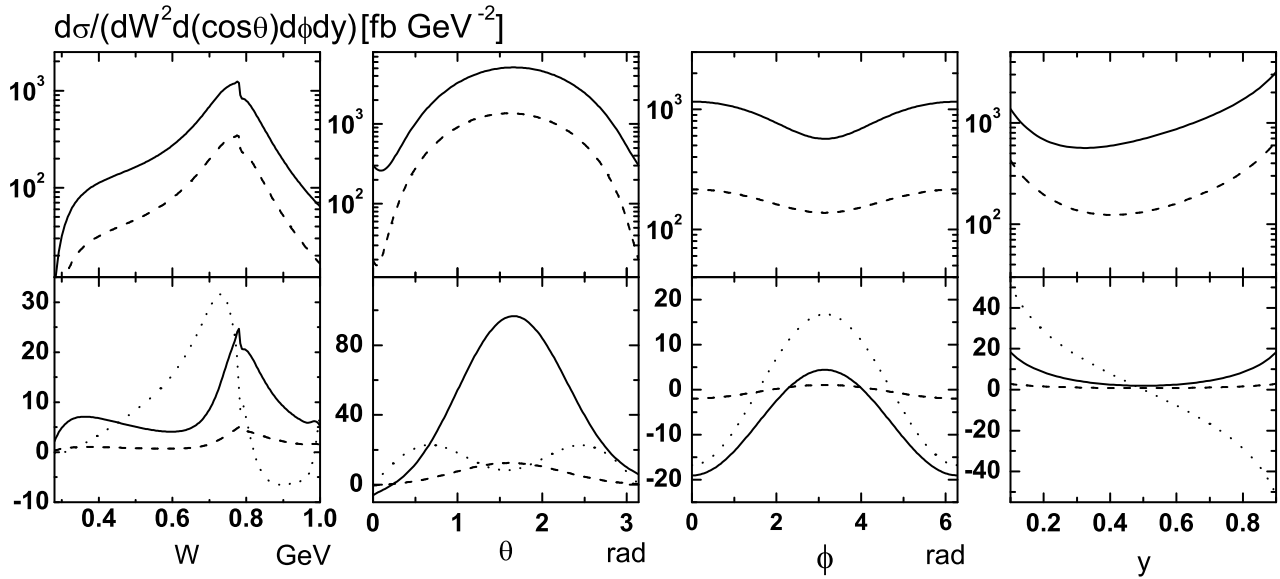}} \caption{\small Upper
panel: the differential cross-sections of the Bremsstrahlung
process, at  $\sqrt{s}=5$ (solid line) and 10 GeV (dashed line).
Lower panel: contributions to the differential cross-sections coming
from the unpolarized interference term between the Bremsstrahlung
process and the process $e^+e^-\rightarrow \gamma^* \rightarrow
\pi^+\pi^-\gamma$ at $\sqrt{s}=5$ (solid line), 10 GeV (dashed line)
and helicity dependent interference term (dotted line) at
$\sqrt{s}=5$ GeV. The first column: the $W$ dependent differential
cross-sections at $\theta=20^\circ$, $\phi=180^\circ$, $y=0.3$; the
second column: the $\theta$ dependent differential cross-sections at
$W=0.7 \,\textrm{GeV}$, $\phi=180^\circ$, $y=0.3$; the third column:
the $\phi$ dependent differential cross-sections at $W=0.7
\,\textrm{GeV}$, $\theta=20^\circ$, $y=0.3$; The fourth column: the
$y$ dependent differential cross-sections at $W=0.7 \,\textrm{GeV}$,
$\theta=20^\circ$, $\phi=180^\circ$.} \label{bremint}
\end{center}

\end{figure*}

The $\gamma Z$ interference term is also interesting because it
contains the weak mixing angle $\sin\theta_W$. To isolate this term
one needs to go beyond the helicity-independent cross-section, that
is, to consider the helicity dependent cross-section of the process.
The helicity dependent differential cross-section is dominated by
the $\gamma Z$ interference term:
\begin{eqnarray}
&&\frac{d\sigma(+\lambda_e)-d\sigma(-\lambda_e)}{2dW^2 d\cos
\theta}\nonumber\\
&=&-\frac{\alpha^3 \beta (s-W^2)}{6 s^3}\chi_{\gamma
Z}c_A^l\lambda_e\nonumber\\
&&\times(e_uc_V^u+e_dc_V^d)(e_u^2+e_d^2)|V_u(\cos\theta,W^2)|^2,
\end{eqnarray}
since the  $Z$ boson term can be ignored compared to that of the
$\gamma Z$ interference term.

 As discussed previously, the unpolarized differential
cross-section is dominated by the virtual photon production, which
reads:
\begin{eqnarray}
&&\frac{d\sigma(+\lambda_e)+d\sigma(-\lambda_e)}{2dW^2 d(\cos
\theta)}\nonumber\\
&=&\frac{\alpha^3 \beta (s-W^2)}{12
s^3}(e_u^2+e_d^2)^2|V_u(\cos\theta,W^2)|^2.
\end{eqnarray}
The $W$-dependence of this cross-ection is shown in
Fig.~\ref{neutralw}, for $\sqrt{s}=10$ GeV and $\theta=20^\circ$.

The ratio of the helicity dependent and helicity-independent
cross-section, that is, the helicity asymmetry of the process is
($\lambda_e=+1$)
\begin{eqnarray}
R&=&\frac{\sigma(+\lambda_e)-\sigma(-\lambda_e)}{\sigma(+\lambda_e)+\sigma(-\lambda_e)}\nonumber\\
&=&-2c_A^l\chi_{\gamma
Z}\frac{e_uc_V^u+e_dc_V^d}{e_u^2+e_d^2}\nonumber\\
&=&\frac{\frac{9}{10}-2\sin^2
\theta_W}{\sin^2(2\theta_W)}\frac{s}{s-M_Z^2},\label{ratio}
\end{eqnarray}
which depends on $s$ and $\sin\theta_W$. Therefore the measurement
on this helicity asymmetry provides an alternative approach to
extract the value of weak mixing angle $\sin\theta_W$. In
Fig.~\ref{rs} we show the $\sqrt{s}$ dependent ratio $R$ which is
given in (\ref{ratio}). For $\sqrt{s}=10\,\textrm{GeV}$, which is
the typical c.m. energy of CLEO or $B$-factories, such as BaBar and
Belle, this ratio is about 0.8\%, which is a sizeable value.

\subsection{Results for the charged case}

In the case of charged pion pair production, besides the process
$e^+e^- \rightarrow \gamma^* \rightarrow \pi^+ \pi^- \gamma$, the
Bremsstrahlung process $e^+e^- \rightarrow \gamma^* \gamma
\rightarrow \pi^+ \pi^- \gamma$ also needs to be considered. The
cross-section of the former process is the same as that of neutral
pion production, since only $\Phi_q^+$ contributes. The differential
cross-section of the Bremsstrahlung process has the form
\begin{eqnarray}
\frac{d \sigma^B }{ d W^2 d ( \cos \theta) d\phi dy}
        &=& \frac{ \alpha^3 \beta^3 (s - W^2) }{ 16 \pi s^2 W^2}
        |F_\pi(W^2)|^2 \nonumber\\
        &&\hspace{-2cm}\times( (1-2x(1-x)) \frac{I_1 (y)}{y(1-y)}\sin^2 \theta\nonumber\\
        &&\hspace{-2cm}+4x(1-x)\cos^2
        \theta\nonumber\\
        &&\hspace{-2cm}+ \sqrt{ x ( 1 - x ) }
        ( 1 - 2 x ) \frac{ I_2 (y) }{ \sqrt{ y ( 1 - y ) } } \sin 2
        \theta \cos \phi \nonumber\\
        &&\hspace{-2 cm}- x (1-x) 2  \sin^2 \theta \cos 2 \phi)
        ,\label{bremcs}
\end{eqnarray}
where $F_\pi(W^2)$ is the timelike pion form factor.

It is interesting to investigate the interference
\cite{brodskylecture,ls05} between $e^+e^- \rightarrow \gamma^*
\rightarrow \pi^+ \pi^- \gamma$ and the Bremsstrahlung process,
because it can be applied to study the subprocess $\gamma^*
\rightarrow \pi^+ \pi^- \gamma$ at the amplitude level, similar to
the case $e\gamma\rightarrow e \pi \pi$ that has been discussed in
Refs.~\cite{Diehl00,lps06}. Knowledge of the process $\gamma^*
\rightarrow \pi^+ \pi^- \gamma$ at the amplitude level can help to
determine not only the magnitude of $\Phi^+_q$, but also its phase.
The form of the interference term reads
\begin{eqnarray}
\frac{d\sigma^I}{dW^2 d(\cos \theta) d\phi dy}&=&
            \frac{\alpha^3\beta^2(s - W^2)}{16\pi s^2\sqrt{s}W}\nonumber\\
            &&\hspace{-2.5cm}\times(\sqrt{x(1-x)}
            \textrm{Re}\{F_\pi^* V\}\cos \theta \nonumber\\
            &&\hspace{-2.5cm}+[\frac{xI_1 (y)}{\sqrt{y(1-y)}}-(1-x)\sqrt{y(1-y)}]\nonumber\\
            &&\hspace{-2.5cm}\times\textrm{Re}\{F_\pi^* V\}\sin\theta \cos \phi \nonumber\\
            &&\hspace{-2.5cm}+\lambda_e
            x\frac{I_2 (y)}{2\sqrt{y(1-y)}}\textrm{Im}\{F_\pi^* V\} \sin \theta \sin
            \phi),\label{intf}
\end{eqnarray}
where $V=\sum_{q} V_q$ for $q=u,d$. This interference term can be
obtained from taking the difference of the cross-section in the
exchange $(\theta, \phi)\rightarrow (\pi-\theta, \pi+\phi)$. The
last term in Eq.~(\ref{intf}) is the helicity dependent term, which
can probe $\textrm{Im}\{F^* V \}$, and which together with the
helicity-independent terms can be used to reconstruct the full
complex amplitude of $\gamma^*\rightarrow \pi^+\pi^-\gamma$, in the
region where $F_\pi(W^2)$ is known. Conversely, these measurements
could also be used in order to obtain information about the pion
timelike form factor $F_\pi(W^2)$.

In Fig.~\ref{bremint} we give the curves of the differential
cross-sections, coming from the Bremsstrahlung process and its
interference with the process $e^+e^- \rightarrow \gamma^*
\rightarrow \pi^+ \pi^- \gamma$, respectively. For $F_\pi(W^2)$ we
use the parametrization $N=1$ given in Ref.~\cite{ks90}. We show the
cross-sections for $\sqrt{s}= 5$ (solid line) and 10 GeV (dashed
line) in Fig.~\ref{bremint}. In the later case the terms containing
$1-x$ have sizeable contributions. In the lower panel of
Fig.~\ref{bremint} we also show the contribution from the helicity
dependent interference term (dotted line) for $\sqrt{s}=5$ GeV.

Fig.~\ref{bremint} shows that the contribution coming from the
Bremsstrahlung process is $1 \sim 2$ orders of magnitude larger than
that from the interference term. According to the shape of $F_\pi$,
the $W$-dependent curves of both the Bremsstrahlung and interference
contributions have peaks in the region $W\sim m_\rho$. In the case
of $\theta$-dependence the curves of Bremsstrahlung and
(unpolarized) interference contributions increase with $\theta$ from
$0^\circ$ to $90^\circ$, and then decrease. The $\phi$-dependent
curve of the Bremsstrahlung contribution has a large value at
$\phi=180^\circ$ and a small value at $\phi=0^\circ$, while that of
the (unpolarized) interference contribution has a large (absolute)
value at $\phi=0^\circ$ and a small (absolute) value at
$\phi=180^\circ$. The $y$-dependence of the interference
contribution shows that the curves have maximum magnitude in the
region $y\rightarrow 0$ or 1 (due to the factor $\sqrt{y(1-y)}$ in
the denominators of some terms in (\ref{intf})), which is therefore
the suitable region in order to study the interference term. A
similar situation occurs in the case of the Bremsstrahlung process.

\section{Summary}

We have given a detailed analysis of the process $e^+e^-\rightarrow
\gamma^*\,\textrm{or} \, Z^*\rightarrow\pi\pi\gamma$, in the
kinematical region where $\sqrt{s}$ is large compared to the
invariant mass of the pion pair but much below the $Z$-pole. A
factorization similar to which has been shown in Ref.~\cite{ls05}
holds for the $\gamma$ case: the amplitude of the process can be
expressed as the convolution between the hard scattering $\gamma^*
\rightarrow q \bar{q} \gamma$ and 2-pion GDAs which describe the
soft transition $q \bar{q} \rightarrow \pi \pi$. Thus the
investigation on the process $e^+e^-\rightarrow
\gamma^*\rightarrow\pi\pi\gamma$ can provide detailed information
about 2-pion GDAs, especially their $C$-even part. The experimental
measurement of this process can also provide experimental indication
of pion GPDs, by virtue of the fact that this process is the crossed
channel of the DVCS process $e\pi \rightarrow e \pi \gamma$, and
there is a crossing relation between GPD and GDA. We studied the
production of both neutral and charged pion. In the former case the
$\gamma Z$ interference term can be applied to extract the value of
the weak mixing angle $\sin \theta_W$ through the measurement of the
helicity asymmetry in the process $e^+e^-\rightarrow
\pi^0\pi^0\gamma$. In the last case the interference between the
Bremsstrahlung process and the process $e^+e^-\rightarrow
\gamma^*\rightarrow \pi^+\pi^-\gamma$ is useful in order to study
the subprocess $\gamma^*\rightarrow \pi^+\pi^-\gamma$ at the
amplitude level.

\begin{acknowledgments}
This work is supported by Fondecyt (Chile) under Projects
No.~3050047 and No.~1030355.
\end{acknowledgments}

\end{document}